\documentclass[aps,prl,twocolumn,groupedaddress,showpacs]{revtex4}

\usepackage{graphicx}
\usepackage{amsmath}

\begin{document}


\title{Uniformly frustrated bosonic Josephson-junction arrays}


\author{Kenichi Kasamatsu}

\affiliation{Department of Physics, Kinki University, Higashi-Osaka, 577-8502, Japan}


\date{\today}

\begin{abstract}
We derive a uniformly frustrated $XY$ model that describes two-dimensional Josephson-junction arrays consisting of rotating Bose-Einstein condensates trapped by both a harmonic trap and a corotating deep optical lattice. The harmonic trap makes the coupling constant of the model have a nonuniform parabolic dependance. We study the ground state through Monte Carlo simulations in a wide range of the frustration parameter $f$, revealing a rich variety of vortex patterns.   
\end{abstract}


\maketitle

Josephson-junction arrays (JJAs), a network of superconducting islands, have attracted much interest because they are well-controlled systems to study nontrivial phase transitions as well as macroscopic quantum phase coherence \cite{Fazio}. The application of transverse magnetic fields to the superconducting JJA leads to realization of the uniformly frustrated $XY$ model (UFXYM) 
\begin{equation}
H = -J \sum_{\langle jj' \rangle} \cos (\theta_{j} - \theta_{j'} +A_{jj'} ). \label{UFXYM}
\end{equation}
Here, $J>0$ denotes the coupling constant, $\theta_{j}$ the phase of the superconducting node at a site $j$, and $\langle jj' \rangle$ near neighbors. The bond variables $A_{jj'}$ satisfy the constraint $\sum A_{jj'} = 2 \pi f$, where the summation is taken over the perimeter of a plaquette of the junctions and $f$ is the magnetic flux (vortex) piecing the plaquette in units of the flux quantum. The vortices induces the frustration for the stable direction of the order parameter's phase at each site. The competition of two length scales --- the mean separation of vortices and the period of underlying lattice --- yields a rich variety of ground state structures, which depend on the rational or irrational number of $f$ \cite{Teitel,Halsey1,Straley,Lee,Halsey}. Also, the nature of the finite-temperature phase transition for nonzero $f$ is still not fully elucidated, while for $f=0$ it is interpreted as the Berezinskii-Kosterlitz-Thouless (BKT) mechanism. For $f=1/2$, in particular, it remains controversial whether there are two distinct phase transitions associated with breaking of the continuous symmetry of $U(1)$ gauge and discrete symmetry of $Z_{2}$ chirality, closely connected with unbinding of kink-antikink pair excitation at Ising-type domain boundaries \cite{Santiago}.

Cold atoms in a optical lattice (OL) provide an ideal testing ground for the study of many-body physics associated with the model Hamiltonian in condensed matter systems \cite{OLreview}. The advantage is that the microscopic parameters of the periodic potential can be precisely controlled. The cold-atom analogs of JJAs have been realized in a one-dimensional (1D) OL \cite{Anderson,Cataliotti}, where many Bose-Einstein condensates (BECs) are separated by potential barriers along the lattice direction. Also, it has suggested that BECs confined by a 2D OL can mimic the physics of 2D JJAs \cite{Smerzi}. Recently, thermally activated vortex formation, associated with the BKT mechanism, in such a 2D bosonic JJA was observed through the direct imaging of the density profile \cite{Schweikhard}. 

In this work, we investigate the rotation effect, analogous to that of a magnetic field for superconductors, on the 2D JJAs consisting of an atomic BEC. A recent experiment by Tung {\it et al.} demonstrated periodic pinning effects for vortices in a BEC by the rotating OL \cite{Tung}. Several theories suggested rich phase diagrams of vortex states due to the interplay between the vortex-vortex interaction and the periodic pinning potential \cite{Reijnders,Pu,Kasamatsu,Sato}. However, they considered them only for a few values of the filling factor, the vortex number per unit cell of the OL (frustration parameter $f$). Here, we consider BECs in a 2D deep OL, where the condensate fractions are well localized at the periodic potential minima to form a 2D JJA. The application of rotation to this system realizes the {\it uniformly frustrated bosonic JJA} \cite{Polini}. The mapping into the UFXYM is helpful to study the equilibrium vortex structure in a wide range of rotation frequency, because direct simulation of the Gross-Pitaevskii equation is a time-consuming work. Also, the model provide simple approach to explore finite-temperature effects, which could provides a new ground to verify unresolved problems in statistical physics described above. In this paper, we clarify the equilibrium vortex configuration in the rotating bosonic JJA using Monte Carlo simulations of the UFXYM in a wide range of the frustration parameter $f$. Since we treat explicitly the trapping potential in addition to the OL, the site-site couplings become nonuniform and a finite-size effect is expected. 

First, we derive the UFXYM to describe the rotating bosonic JJA combined with the harmonic trap. The BECs in a deep 2D OL can be mapped onto the $XY$ model, where the amplitude of the condensate wave function is frozen at each site, but its phase is still a relevant variable \cite{Smerzi}. Here, we make use of this formalism for the rotating system. The many-body Hamiltonian of bosons in a rotating frame with frequency ${\bf \Omega} = \Omega \hat{\bf z}$ is 
\begin{equation}
\hat{H} = \int d {\bf r} \hat{\psi}^{\dagger} \left[ \frac{( - i \hbar \nabla - m {\bf \Omega} \times {\bf r} )^{2}}{2m} + V_{\rm ex} +\frac{g}{2} \hat{\psi}^{\dagger} \hat{\psi} - \mu \right] \hat{\psi}, \label{hamiltonian}
\end{equation}
where $m$ is the atomic mass and $g = 4 \pi \hbar^{2} a / m$ the coupling constant with $s$-wave scattering length $a$. The field operator $\hat{\psi}$ obeys the bosonic commutation relations. Conservation of the total particle number is ensured by the chemical potential $\mu$. The external potential consists of two parts $V_{\rm ex}=V_{\rm ho} + V_{\rm OL}$: a centrifugal-force-modified harmonic potential $V_{\rm ho} = m (\omega_{\perp}^{2} - \Omega^{2}) r^{2} / 2 + m \omega_{z}^{2} z^{2} / 2 $ and a 2D OL $V_{\rm OL} = V_{0} [\sin^{2} (\pi x/d) + \sin^{2} (\pi y/d) ]$ with the square lattice geometry and the spatial periodicity $d$. The minima of the 2D OL are located at the points ${\bf j} d = (j_{x}, j_{y}) d$ with integers $j_{x}$ and $j_{y}$. 

We assume that the laser intensity is large enough to create many separated wells giving rise to a 2D array of condensates. Still, the small overlap between the wave functions of adjacent wells causes quantum tunneling and can be sufficient to ensure overall coherence of the system. If the energy due to interaction and rotation is small compared to the energy separation between the lowest and first excited band, the particles are confined to the lowest Wannier orbitals. Following the analogy of a Bloch electron in a magnetic field, we take the Wannier basis as $\hat{\psi}({\bf r}) = \sum_{\bf j} \hat{a}_{\bf j} w_{\bf j}({\bf r}) \exp \left[ (im/\hbar) \int_{{\bf r}_{\bf j}}^{\bf r} {\bf A}({\bf r}') \cdot d {\bf r}' \right]$, where ${\bf A} = {\bf \Omega} \times {\bf r}$ is the analog of the magnetic vector potential, $w_{\bf j}({\bf r})$ the Wannier wave function localized at the ${\bf j}$th well, and $\hat{a}_{\bf j}$ the boson annihilation operator. The normalization condition $\int d {\bf r} w^{\ast}_{\bf j}({\bf r}) w_{{\bf j}'}({\bf r}) = \delta_{{\bf j},{\bf j}'}$ implies the total number $N=\sum_{\bf j} \langle \hat{a}^{\dag}_{\bf j} \hat{a}_{\bf j} \rangle \equiv \sum_{\bf j} N_{\bf j}$.

With this basis, Eq. (\ref{hamiltonian}) leads to the Bose-Hubbard model in the rotating frame \cite{Wu}
\begin{eqnarray} 
\hat{H}= -\sum_{\langle {\bf j} , {\bf j}' \rangle}  \frac{t_{{\bf j} , {\bf j}'}}{2} (\hat{a}^{\dag}_{\bf j} \hat{a}_{{\bf j}'} e^{-i A_{{\bf j} , {\bf j}'}} +  {\rm h.c.}) +\sum_{\bf j} E_{\bf j} \hat{N}_{\bf j} \nonumber \\ 
+ \sum_{\bf j} \frac{U_{\bf j}}{2} \hat{N}_{\bf j}  (\hat{N}_{\bf j} - 1) , \label{BHM} 
\end{eqnarray} 
where $\sum_{\langle {\bf j}, {\bf j}' \rangle}$ denotes a sum over nearest-neighbor sites and $t_{{\bf j} , {\bf j}'} = - \int d {\bf r} w_{\bf j}^{\ast}({\bf r}) \left( - \hbar^2 \nabla^{2}/2m  + V_{\rm OL} \right) w_{{\bf j}'}({\bf r}) $, $E_{\bf j} = \int d {\bf r} w_{\bf j}^{\ast}({\bf r})  \left( - \hbar^2\nabla^{2} /2m + V_{\rm ex} - \mu \right)  w_{\bf j}({\bf r}) $, and $U_{\bf j} = g \int d {\bf r} |w_{\bf j}({\bf r}) |^4$ represent the hopping matrix element, the energy offset of each lattice site, and the on-site energy, respectively. The effect of rotation is described by $A_{{\bf j},{\bf j}'} = (m/\hbar) \int_{{\bf r}_{\bf j}}^{{\bf r}_{{\bf j}'}} {\bf A}({\bf r}') \cdot d{\bf r}' $ with the constraint $\sum_{u.c.} A_{{\bf j}, {\bf j}'} = 2 \pi f$, where the sum is taken around any unit cell of the 2D array. The constant $f$ is the frustration parameter, being given by the average number of vortices per unit cell: $f = 2 \Omega d^{2} / \kappa$, with quantum circulation $\kappa = h/m$. The Hamiltonian ($\ref{BHM}$) predicts novel vortex properties and fractal quantum Hall features of the strongly interacting lattice bosons \cite{Wu,Sorensen}. Other methods of creating this ``effective" magnetic field have been discussed \cite{Jaksch}.

If the number of atoms per site is large ($N_{\bf j} \gg 1$), the operator can be expressed in terms of its amplitude and phase, the amplitude being subsequently approximated by the $c$ number as $\hat{a}_{\bf j} \simeq \sqrt{N_{\bf j}} e^{i \hat{\theta}_{\bf j}}$. Then, Eq. (\ref{BHM}) reduces to 
\begin{eqnarray} 
\hat{H}= -  \sum_{\langle {\bf j},{\bf j}' \rangle} J_{{\bf j},{\bf j}'} \cos{(\theta_{\bf j} - \theta_{{\bf j}'} + A_{{\bf j},{\bf j}'})} - \sum_{\bf j} {U_{\bf j} \over 2} \frac{\partial^2}{\partial \theta_{\bf j}^2} \nonumber \\ 
- i \sum_{\bf j} (E_{\bf j} +U_{\bf j} N_{\bf j}) \frac{\partial}{\partial\theta_{\bf j}} + \sum_{\bf j} \left( E_{\bf j} N_{\bf j} + \frac{U_{\bf j}}{2} N_{\bf j}^{2} \right), 
\label{QPM} 
\end{eqnarray}  
where we have used the phase representation $\hat{N}_{\bf j} = N_{\bf j} - i \partial/\partial \theta_{\bf j}$, $\hat{\theta}_{\bf j} = \theta_{\bf j}$, and the notation $J_{{\bf j},{\bf j}'} = t_{{\bf j}, {\bf j}'} \sqrt{N_{\bf j} N_{{\bf j}'}}$. This reduction is valid when $J_{{\bf j},{\bf j}'}/N_{\bf j}^{2} \ll U_{\bf j}$ \cite{Smerzi}. 

The first term of Eq. (\ref{QPM}) corresponds to the UFXYM with spatially inhomogeneous nearest-neighbor coupling $J_{{\bf j}, {\bf j}'}$. To neglect the other terms and to estimate $J_{{\bf j}, {\bf j}'}$, the equilibrium form of $w_{\bf j}$ and $N_{\bf j}$ must be calculated. We assume that the equilibrium density is determined by minimizing the last $c$ number term of Eq. (\ref{QPM}), which is the dominant contribution of the ground-state energy. Then, $E_{\bf j} +U_{\bf j} N_{\bf j} = 0$ and the third term may be neglected automatically. Next, we apply the ansatz $w_{\bf j}({\bf r}) = u_{\bf 0} (x-j_{x}d,y-j_{y}d) v_{\bf j}(z)$ with the site-{\it independent} transverse part $u_{\bf 0} (x,y) = (\sqrt{\pi} \sigma)^{-1} e^{-(x^{2}+y^{2})/2 \sigma^{2}}$ and the site-{\it dependent} longitudinal part $v_{\bf j}(z)$ \cite{Kramer}. Since the atoms are tightly confined by 2D OL, the contribution arising from the two-body interactions is negligible for the estimation of $u_{\bf 0}(x,y)$ and the the variational parameter $\sigma$ can be obtained easily. The longitudinal part is approximated by the inverted parabolic form $v_{\bf j} (z)^{2} = (\mu_{\bf j}/g_{\rm 1D} N_{\bf j}) ( 1 - z^{2}/R_{z {\bf j}}^{2} )$, with $g_{\rm 1D} = g/2 \pi \sigma^{2}$, the local chemical potential $\mu_{\bf j} =  m (\omega^{2} - \Omega^{2}) (j_{\rm max}^{2} - j_{x}^{2} - j_{y}^{2}) d^{2}/2$, and the Thomas-Fermi radius $R_{z {\bf j}}^{2} = 2 \mu_{\bf j}/m\omega_{z}^{2}$. Here, $N_{\bf j} = 0$ for $|{\bf j}| > j_{\rm max}$ because of the harmonic confinement. Using the normalization condition $\int v_{\bf j} (z)^{2} dz = 1$ and $\sum_{\bf j} N_{\bf j} = N$, we can obtain 
\begin{equation}
j_{\rm max} = \frac{a_{\perp}}{d} \left( \frac{15N}{2\pi} \frac{\omega_{z}}{\omega_{\perp}} \frac{a d^{2}}{a_{\perp} \sigma^{2}}\right)^{1/5} \left( 1- \frac{\Omega^{2}}{\omega_{\perp}^{2}} \right)^{-3/10} \label{jmaxmax}
\end{equation}
with $a_{\perp} = \sqrt{\hbar/m \omega_{\perp}}$ and 
\begin{equation}
N_{\bf j} = \frac{5 N}{2 \pi j_{\rm max}^{2}} \left( 1-\frac{j_{x}^{2}+j_{y}^{2}}{j_{\rm max}^{2}} \right)^{3/2}. \label{particledist}
\end{equation}
For a given $V_{0}$ we evaluate the variational wave function $u_{\bf 0} (x,y)$ to obtain the optimized value of $\sigma$. Through Eqs. (\ref{jmaxmax}) and (\ref{particledist}) with this optimized $\sigma$, the parameter values in Eq. (\ref{QPM}) as well as $N_{\bf j}$ can be fixed. 

Under these formula we investigate the ground state of this system. Following the typical experimental conditions such as $^{87}$Rb atoms used in JILA experiments \cite{Tung,Schweikhard}, we use $N = 6 \times 10^{5}$ and $a=5.29$ nm. The frequencies of the trapping potential are set as $\omega_{\perp} = 11.5 \times 2\pi$ and $\omega_{z} = 50 \times 2\pi$, which gives $a_{\perp} =$3.2 $\mu$m. The lattice spacing is set as $d = 5$ $\mu$m. 

We confirm that the obtained distribution $N_{\bf j}$ is quantitatively consistent with that obtained from the numerical solution of the 3D Gross-Pitaevskii equation; the particle number at the central well is $N_{(0,0)} \simeq 6000 $, decreasing from the center to the outside according to Eq. (\ref{particledist}). The conditions of the Josephson regime, $J_{{\bf j},{\bf j}'}/N_{\bf j}^{2} \ll U_{\bf j}$ and $J_{{\bf j},{\bf j}'} \gg U_{\bf j}$, are certainly satisfied. The former condition is valid because of $N_{\bf j} \gg 1$, even for outermost sites with $N_{\bf j} \sim 100$. For the central region $(j_{x}, j_{y})=(0,0)$ $(j_{x}', j_{y}')=(1,0)$, the condition $J_{{\bf j},{\bf j}'} \gg U_{\bf j}$ is well satisfied for $30 \leq V_{0}/\hbar \omega_{\perp} \leq 90$. We take $V_{0} = 65 \hbar \omega_{\perp} $ in the following discussion, having $J_{(0,0),(1,0)} /  U_{(0,0)} \simeq 100$ and $J_{(0,0),(1,0)} = 0.9025 \hbar \omega_{\perp}$. Even for $|{\bf j}| \simeq j_{\rm max}$, the condition $J_{{\bf j},{\bf j}'} \gg U_{\bf j}$ is still good. Therefore, the quantum correction arising from the third term of Eq. (\ref{QPM}) may be neglected in our problem. 

We perform Monte Carlo simulations of the Hamiltonian 
\begin{equation}
H = - \sum_{\langle {\bf j}, {\bf j}' \rangle} J_{{\bf j}, {\bf j}'} 
\cos (\theta_{\bf j} - \theta_{{\bf j}'} +A_{{\bf j}, {\bf j}'} ). \label{UFXYMbec}
\end{equation}
The form of the coupling energy is 
\begin{equation}
J_{{\bf j},{\bf j}'} \simeq \sqrt{N_{\bf j} N_{{\bf j}'}} e^{-d^{2}/4\sigma^{2}}  \biggl[ \frac{\hbar^{2}}{2 m \sigma^{2}} \left(\frac{d^{2}}{4 \sigma^{2}} -1 \right) - V_{0}  \biggr], 
\end{equation}
where we have used the optimized value of $\sigma$ and, when calculating the integral in $J_{{\bf j},{\bf j}'} $, the integral for the $z$ direction was approximated as $\int_{-R_{z {\bf j}'}}^{R_{z {\bf j}'}} dz v_{\bf j} v_{{\bf j}'} \simeq \sqrt{\int_{-R_{z {\bf j}}}^{R_{z {\bf j}}} dz v_{\bf j}^{2} \int_{-R_{z {\bf j}'}}^{R_{z {\bf j}'}} dz v_{{\bf j}'}^{2} }$ with Thomas-Fermi radius $R_{z {\bf j}} \geq  R_{z {{\bf j}'}}$ and the area of the integral for the $xy$ plane as $\int_{0}^{d} dx \int_{-\infty}^{+ \infty} dy u_{\bf 0}(x,y) V_{\rm OL} u_{\bf 0}(x,y)$. The symmetric gauge is chosen for the vector potential $A_{{\bf j}, {\bf j}'}$. We use the Metropolis algorithm to study the ground-state properties of this system as a function of the frustration parameter $f$. For this purpose, the temperature is gradually decreased from high temperatures to zero according to the stimulated annealing. Since there are many metastable state caused by the frustration, we change the annealing rates in the several hundred simulations, taking the steady solution with the lowest energy as the ground state. 

It is known that the UFXYM of Eq. (\ref{UFXYM}) exhibits rich ground state structures depending on the parameter $f$ \cite{Teitel,Halsey1,Straley}. For rational $f=p/q$, the ground state is periodic on the $q \times q$ cell in most cases. The striking difference of Eqs. (\ref{UFXYM}) and (\ref{QPM}) of the bosonic JJA is the inhomogeneous coupling $J_{{\bf j},{\bf j}'} \propto \sqrt{N_{\bf j} N_{{\bf j}'}} $. Also, it should be noted that the range of $f$ is restricted by the harmonic potential because the rotation frequency $\Omega$ cannot exceed $\omega_{\perp}$ --- that is, $f < d^{2}/\pi a_{\perp}^{2} = 0.78$ in our case.  

\begin{figure}
\includegraphics[height=0.42\textheight]{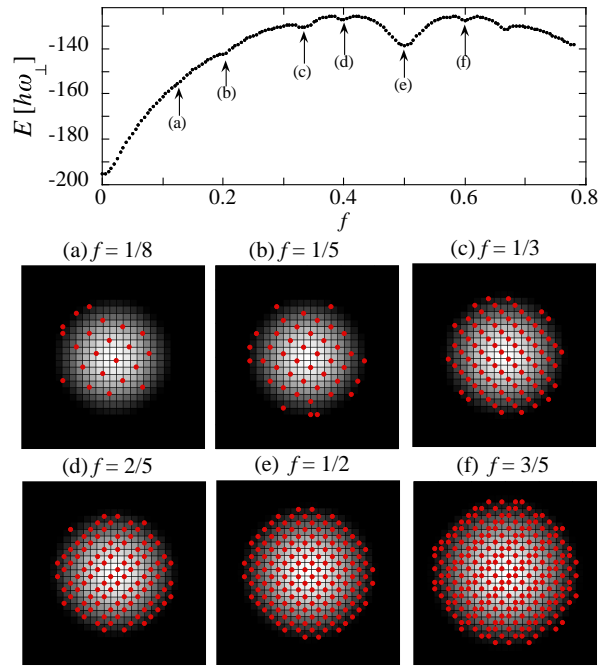}%
\caption{(Color online) Ground-state energy and vortex lattice structures in a bosonic JJA under rotation. The top panel shows the total energy (normalized by $J_{(0,0),(1,0)}$ for $f=0$) as a function of $f$. The bottom panels from (a) to (f) represent  the discretized condensate density $N_{{\bf j}, {\bf j}'}$ (black-white contour plot) and the positions of vortices marked by gray or red circles. Each square in the density corresponds to the site (minima of the OL), and vortices are located at the corners of the squares (maxima of the OL). The positions of vortices are calculated by the current circulation $\sum \sin (\theta_{\bf j} - \theta_{{\bf j}'} + A_{{\bf j},{\bf j}'})$ with the plaquette sum. The parameter values used are $N=6 \times 10^{5}$, $a=5.29$ nm, $\omega_{\perp}=11.5 \times 2\pi$ Hz, $\omega_{z}=50 \times 2\pi$ Hz, $V_{0}=65\hbar \omega_{\perp}$, and a length of one side of the square, $d=5$ $\mu$m. }
\label{density}
\end{figure}
Figure \ref{density} represents the total energy and the typical vortex patterns of the ground state as a function of $f$. The energy curve has a nonmonotonic behavior characterized by some minima at the simple rational values. These features are reflected in the bottom edge of Hofstadter butterfly spectrum \cite{Straley}. The vortex configurations at these minima possess simple periodic structures as shown in Figs. \ref{density}(a)-(f), which represent the ground state for several values of $f$ giving the visible minima of the energy curve. The vortices form a Bravais lattice with a unit cell of $q \times q$ and a quasi-1D structure oriented in parallel with one of the diagonals of the square lattice \cite{Halsey1,Straley,Lee}. This structure, called staircase states where constant currents flow along the diagonal staircases, was shown to be the true ground state for some limited values of $f$ with simple rational forms such as $f=$1/2, 1/3, 2/5, 3/7, 3/8 in the UFXYM with homogeneous coupling $J$ \cite{Halsey1}. While the periodicity of the vortex positions breaks slightly near the condensate edge, this staircase state can be the ground state for the inhomogeneous trapped system. For $f=1/2$, a fully frustrated case, the vortex lattices form a checkerboard pattern, agreement with the previous studies for trapped BECs \cite{Reijnders,Pu}. The energy is approximately reflection symmetric about $f = 1/2$ \cite{tyuu}, and the periodic structures for $f>1/2$ are equivalent to those of $1-f$, but the condensate size is expanded and vortices are replaced by ``vacancies"; an example is shown in Figs. \ref{density} (d) and \ref{density} (f). 

\begin{figure}
\includegraphics[height=0.25\textheight]{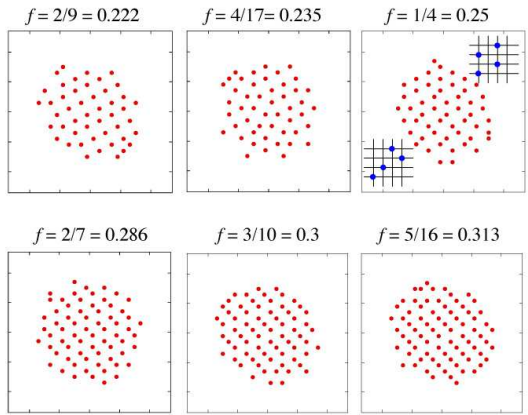}%
\caption{(Color online) The typical intermediate structures between $f=1/5$ and $f=1/3$. For $f = 1/4$ we also show two degenerate unit-cell structures of the ground state for the homogenous system.}
\label{density2}
\end{figure}
Between these energy minima, we obtain characteristic intermediate structures consisting of the domains of simple periodic Bravais lattices; Fig. \ref{density2} shows an example of how one simple periodic structure ($f = 1/5$) changes to another ($f = 1/3$). Since the ground state has typically $q \times q$ periodic unit cells, it is difficult to obtain the periodic structure for large $q$ in the finite-size system. The periodicity is easily broken near the condensate edge due to the weak couplings \cite{Kragset}, the structural change being of a crossover. This is contrast to the homogeneous model where the vortex patterns and accompanying domain walls form diagonal lines for a square lattice, except for irrational values of $f$ \cite{Halsey}. This broken periodicity does not become noticeable as $f$ increases, because the system size expands due to the centrifugal effect and approaches the homogeneous limit. For $1/3< f < 1/2$ the results reproduce the results obtained by the Coulomb gas model \cite{Lee}. They consist of diagonal domains of the $f=1/2$ checkerboard configuration, separated by domain walls (or domains) of $f=1/3$ structure. For $f > 0.425$, the ground-state structures are the $f=1/2$ checkerboard pattern with a low concentration of missing vortices. 

An interesting case is for $f=1/4$, where two possible vortex configurations of the ground state were proposed for the homogenous system as in Fig. \ref{density2} \cite{Teitel,Straley}; these two configurations have exactly the same energy per site, and thus they are both ground states. Our simulations show that these two configurations are always separated by curved domain walls. In contrast, the variational result in Ref. \cite{Reijnders} does not evidence the presence
of degenerate configurations with the same energy. 

In conclusion, we derived a realistic UFXYM that describes rotating BECs in both a trapping potential and a corotating deep OL. Monte Carlo simulations of this model clarify a variety of vortex phases for a wide range of the frustration parameter $f$ that have not been predicted  by the Gross-Pitaevskii model. In future work, we plan to study finite-temperature properties such as an analog of competing phase transitions between the BKT type and the Ising type \cite{Santiago} in this inhomogeneous system.

K.K. acknowledges supports of a Grant-in-Aid for Scientific Research from JSPS (Grant No. 18740213). 



\begin{thebibliography}{99}
\bibitem{tyuu}
Due to the dependence of $J_{{\bf j},{\bf j}'}$ on $f$, such symmetry should not be exact 
as for the homogeneous UFXYM. Since $J_{{\bf j},{\bf j}'} \propto \sqrt{N_{\bf j} N_{{\bf j}'}}$ 
and the total particle number is fixed as $\sum_{\bf j} N_{\bf j} = N$, the energy takes 
similar values for $f$ and $1-f$ as a result of the summation in Eq. (\ref{UFXYMbec}).
\end{thebibliography}
\end{document}